\documentclass[journal]{IEEEtran}

\usepackage[T1]{fontenc}
\usepackage{amsmath,amsfonts}
\usepackage{algorithm}
\usepackage{array}
\usepackage{caption}
\usepackage{subcaption}   

\usepackage{textcomp}
\usepackage{stfloats}
\usepackage{url}
\usepackage{verbatim}
\usepackage{graphicx}
\usepackage{cite}
\usepackage{cases}
\usepackage{algpseudocode}
\usepackage{amssymb}
\usepackage[
    colorlinks=true,
    linkcolor=blue,
    citecolor=blue,
    urlcolor=magenta
]{hyperref}
\usepackage{xcolor}
\usepackage{float}
\usepackage{placeins} 
\usepackage{amsthm}

\def\BibTeX{{\rm B\kern-.05em{\sc i\kern-.025em b}\kern-.08em
    T\kern-.1667em\lower.7ex\hbox{E}\kern-.125emX}}
\usepackage {balance}

\begin{document}

\title{Constrained Tensor Decomposition-Based Target Sensing for Sparse Non-Uniform Array-Enabled AFDM ISAC Systems}

\author{Linchu Chen, Zhendong Li, Lin Chen, Zhou Su, Wenjie Wang and Wen Chen

\thanks{Linchu Chen, Zhendong Li and Wenjie Wang are with the School of Information and Communication Engineering, Xi’an Jiaotong University, Xi’an 710049, China (email: chenlinchu@stu.xjtu.edu.cn; lizhendong@xjtu.edu.cn; wjwang@xjtu.edu.cn). Lin Chen is with the Department of Electrical and Computer Engineering, Stevens Institute of Technology, Hoboken, NJ 07030, USA (e-mail: lchen53@stevens.edu). Zhou Su is with the School of Cyber Science and Engineering, Xi'an Jiaotong University, Xi'an 710049, China (email: zhousu@ieee.org). Wen Chen is with the School of Integrated Circuits, Shanghai Jiao Tong University, Shanghai 200240, China (e-mail: wenchen@sjtu.edu.cn). (Corresponding author: Zhendong Li)
}
\vspace{-0.5cm}
}

\maketitle

\begin{abstract}
Sparse non-uniform array-enabled affine frequency division multiplexing (AFDM) is a promising candidate for integrated sensing and communication (ISAC), while its performance critically depends on accurate target parameter estimation. In this paper, we propose a constrained tensor decomposition-based sensing framework for delay, Doppler, and angle estimation. Specifically, a manifold-constrained alternating least squares algorithm is developed by exploiting the sparse array geometry structure, enabling robust factor matrix extraction and direct angle estimation. In addition, we analyze the uniqueness condition of the tensor decomposition. From the decomposed factor matrices,  delays and Doppler shifts are coarsely estimated by pulse compression,  and refined by an iterative one dimensional golden-section search.Moreover, the Cramér-Rao bound (CRB) is derived to characterize the fundamental performance limit of the considered estimation problem. Finally, simulation results validate the effectiveness of the proposed estimation algorithm.
\end{abstract}


\section{Introduction}

\IEEEPARstart{I}{ntegrated} sensing and communication (ISAC) has emerged as a key enabler for future wireless systems, with particularly important applications in high-mobility scenarios such as vehicular networks\cite{ref1,ref2}. In such environments, doubly dispersive channels with concurrent multipath delay and Doppler spread pose severe challenges. Orthogonal frequency division multiplexing (OFDM) suffers from significant intercarrier interference due to broken subcarrier orthogonality, which drastically degrades both communication and sensing performance\cite{ref3}. These limitations motivate the exploration of novel waveforms resilient to doubly dispersive channels.

Affine frequency division multiplexing (AFDM) has recently been proposed as a chirp-based multicarrier waveform\cite{ref4,ref5}. Unlike OFDM which uses sinusoidal subcarriers, AFDM employs linear chirp signals as orthogonal basis functions via the discrete affine Fourier transform (DAFT). This chirp structure enables explicit decoupling of delay and Doppler shifts in the DAFT domain, where they manifest as positional shifts and phase rotations, respectively. Consequently, AFDM achieves full diversity gain in doubly dispersive channels and allows low-complexity equalization\cite{ref4}. Compared with orthogonal time-frequency space (OTFS) modulation, AFDM attains comparable error performance with significantly lower complexity and pilot overhead\cite{ref5}. Moreover, its chirp basis inherently exhibits radar-like properties similar to frequency-modulated continuous-wave signals, making AFDM particularly attractive for ISAC without requiring dedicated sensing waveforms\cite{ref6}. While AFDM exhibits remarkable adaptability to doubly dispersive channels at the waveform level, its potential for high‑resolution sensing and massive connectivity still requires synergy with spatial‑domain design to be fully unleashed.

 Conventional compact uniform linear arrays with half-wavelength spacing in AFDM ISAC systems face severe bottlenecks as element count grows: excessive hardware cost, power consumption, and mutual coupling \cite{ref8}. Non-uniform sparse antenna arrays, employing aperiodic spacing exceeding half a wavelength, achieve comparable apertures with far fewer elements, delivering superior spatial resolution and angle estimation accuracy while effectively suppressing grating lobes \cite{ref7}. Their integration with AFDM enables natural complementarity between delay-Doppler domain sparsity and array geometry domain sparsity, jointly supporting better communication and sensing under low hardware overhead.

To fully leverage advantages of sparse non‑uniform array enabled AFDM ISAC system, accurate target parameter information is a prerequisite and a critical factor. Specifically, the receiver must estimate delay, Doppler shift, angle-of-arrival (AoA)  and angle-of-departure (AoD). However, these parameters are highly coupled in the high-dimensional received signal \cite{ref9}, causing conventional estimation algorithms to suffer from error accumulation and limited precision. Tensor decomposition provides a powerful framework for addressing this coupled high-dimensional problem. This is because tensors naturally preserve multi-dimensional signal structures, and the CANDECOMP/PARAFAC (CP) decomposition has proven effective in multiple-input multiple-output (MIMO) radar and communication systems \cite{ref10}. Recent studies have investigated tensor decomposition-based parameter estimation for AFDM ISAC. A tensor‑based angle-delay-Doppler estimation scheme for AFDM-ISAC system in hybrid field scenarios is proposed in\cite{ref12}. Nevertheless, existing tensor-based parameter estimation methods are not applicable to sparse non-uniform arrays, as the irregular spatial sampling fundamentally violates the structured array assumptions underlying current tensor models.

Motivated by this, we investigate target sensing parameter estimation in sparse non-uniform array-enabled AFDM ISAC systems via constrained tensor decomposition based on the array geometry and signal structures. First, we formulate the received signal as a high‑order tensor. Subsequently, a constrained alternating least-squares (CALS) algorithm is used to extract the delay, Doppler and angular parameters from the corresponding factor matrices. Finally, theoretical analysis and simulations validate the proposed algorithm advantages.

\section{System Model}
\begin{figure}
    \centering
    \includegraphics[width=\linewidth]{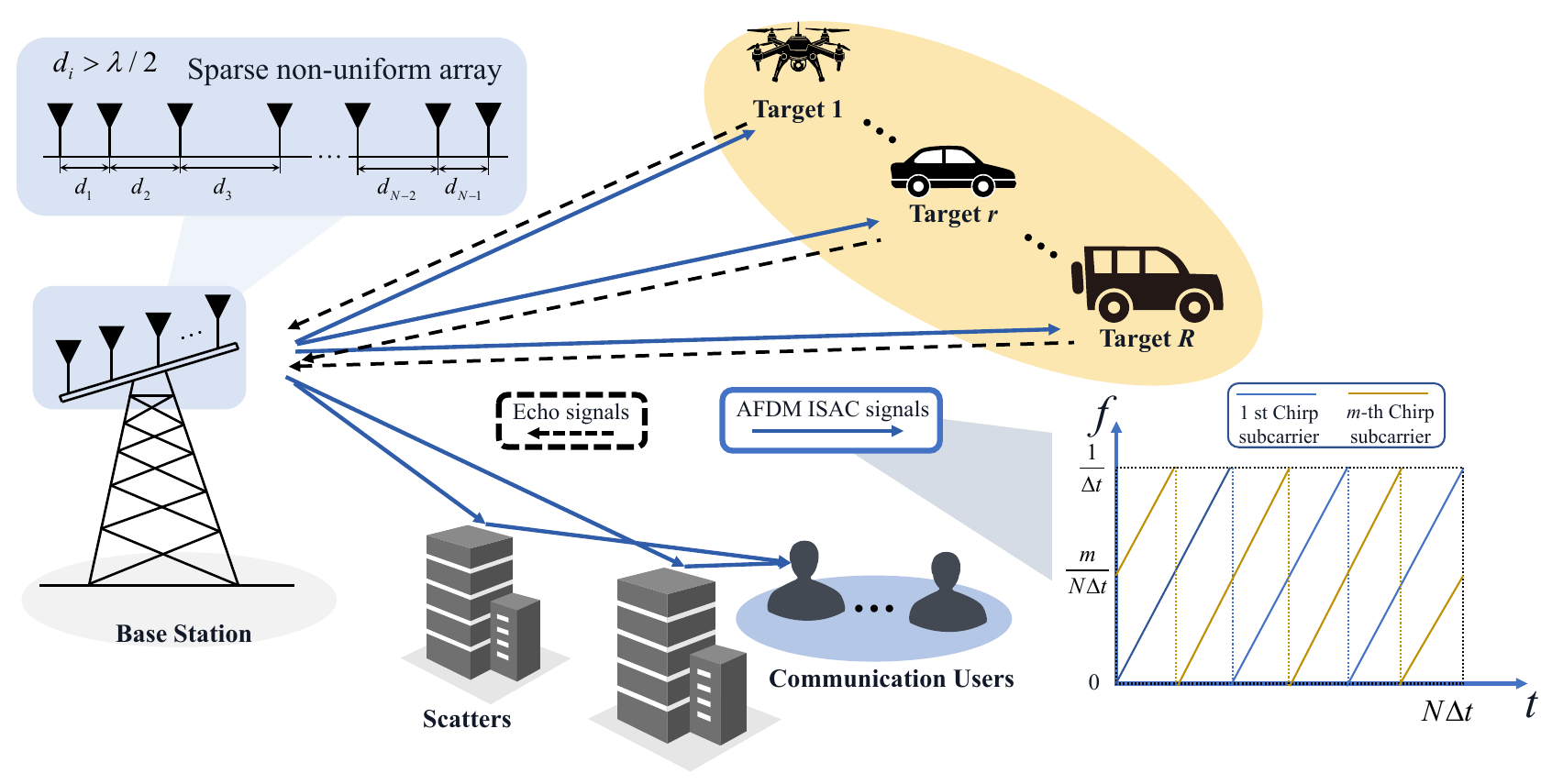}
    \caption{Sparse non-uniform array-enabled AFDM ISAC systems.}
    \label{model}
\end{figure}
As shown in Fig. \ref{model}, we consider a sparse non-uniform array-enabled AFDM ISAC system that performs sensing of $R$ surrounding targets while simultaneously supporting communication with several users in the vicinity of the base station. The base station is equipped with linear sparse non-uniform arrays at both the transmit and receive sides, with $N_t$ and $N_r$ antenna elements, respectively. The system bandwidth $B$ is divided into $N$ chirp subcarriers, with the subcarrier spacing satisfying $\Delta f = B/N$. The sampling interval is $T_s=1/B$, and the symbol duration is $T_a=NT_s$, so that $\Delta f = 1/T_a$. 
Such sparse arrays can achieve sidelobe suppression and other functionalities through specific placement of antenna positions, while realizing a larger array aperture for a given number of antenna elements. To fully characterize the position properties of the sparse non-uniform antenna array, let $\mathbf{t}=[x_1^t,x^t_2,\cdots,x^t_{N_t}]\in\mathbb{R}^{1 \times N_t}$ denote the transmit antenna position coordinates, and $\mathbf{r}=[x_1^r,x^r_2,\cdots,x^r_{N_r}]\in\mathbb{R}^{1 \times N_r}$ denote the receive antenna position coordinates. Then the transmit array steering vector is given by
\begin{equation}
    \mathbf{a}(\mathbf{t};\theta)=[e^{j\frac{2\pi}{\lambda}\rho(x^t_1; \theta)},\cdots,\ e^{j\frac{2\pi}{\lambda}\rho(x^t_{N_t}; \theta)}]^T,
    \label{tdaoxiangshiliang}
\end{equation}
and the receive array steering vector is given by
\begin{equation}
    \mathbf{a}(\mathbf{r};\phi)=[e^{j\frac{2\pi}{\lambda}\rho(x^r_1; \phi)},\cdots,\ e^{j\frac{2\pi}{\lambda}\rho(x^r_{N_r}; \phi)}]^T,
    \label{rdaoxiangshiliang}
\end{equation}
where $\rho(x;\theta)=x\cos\theta$ denotes the path-length difference relative to the reference position of different antennas.

Let $\mathbf{x}\in \mathbb{C}^{N\times 1}$ denote the information symbols in the discrete affine Fourier domain. After AFDM modulation, the transmitted signal is given by $\mathbf{s}[n] = \sum_{m=0}^{N-1}\mathbf{x}[m]\Phi_n[m]$,
where $\Phi_n[m] = \frac{1}{\sqrt{N}}e^{j2\pi(c_1n^2+c_2m^2+mn/N)}$ denotes the affine Fourier basis function, with $c_1$ and $c_2$ being the AFDM parameters. By appropriately setting these two parameters, full diversity gain can be achieved. The matrix representation of the above transform is given by $\mathbf{s} = \mathbf{U}^H \mathbf{x} = \boldsymbol{\Lambda}^H_{c_1} \mathbf{F}^H\boldsymbol{\Lambda}^H_{c_2}\mathbf{x}$, where $\boldsymbol{\Lambda}_c = \operatorname{diag}\{e^{-j2\pi c n^2}\}_{n=0}^{N-1}$, $\mathbf{F}\in\mathbb{C}^{N\times N}$ is the unitary  {discrete Fourier transform (DFT)} matrix with $[\mathbf{F}]_{m,n}=\frac{1}{\sqrt{N}}e^{-j2\pi mn/N}$, and $\mathbf{U}=\boldsymbol{\Lambda}_{c_2}\mathbf{F}\boldsymbol{\Lambda}_{c_1}\in\mathbb{C}^{N\times N}$ is the DAFT matrix. In AFDM ISAC systems, to combat inter-symbol interference caused by multipath delay, a chirp period prefix (CPP) of length $L_\text{CPP}$ can be introduced as $\mathbf{s}[n] = \mathbf{s}[N+n]e^{-j2\pi c_1(N^2+2Nn)}$, where $n =-L_\text{CPP},\cdots,-1$. By exploiting the periodicity of the exponential factor in DAFT, the chirp period prefix is equivalent to the conventional cyclic prefix in OFDM when $2N c_1$ is an integer and $N$ is even.

After serial-to-parallel conversion, the information symbols $\mathbf{x}$ are transmitted from the transmit antenna array and reach the receiver after propagating through the channel. The time-domain channel impulse response is expressed as
\begin{equation}
    h_n[l] = \sum_{r=1}^{R}\beta_r e^{-j2\pi n f^\text{d}_r T_s}\delta[l-\tau_r],
\end{equation}
where $f^\text{d}$ denotes the Doppler shift caused by mobility, and $\tau$ denotes the normalized integer delay $\tau = \overline{\tau}/T_s$, with $L_{\text{CPP}} \geq \max\{\tau_r\}$.  The normalized Doppler frequency is defined as $\nu_r = N f^\text{d}_r T_s = \alpha_r + a_r$, where $\alpha_r=\lfloor\nu_r\rfloor$ and $a_r\in[0,1)$ are the integer and fractional Doppler components.  

After removing the chirp period prefix at the receiver, the matrix representation of the time-domain received signal is given by $\mathbf{r} = \mathbf{H}\mathbf{s} + \mathbf{w}$, where $\mathbf{w}$ denotes the noise vector. The propagation channel $\mathbf{H}$ can be further represented as
\begin{equation}
    \mathbf{H} = \sum_{r=1}^R \beta_r \boldsymbol{\Gamma}_{\text{CPP}_r} \boldsymbol{\Delta}_{f^\text{d}_r} \boldsymbol{\Pi}^{\tau_r},
\end{equation}
where $\boldsymbol{\Pi} \in \mathbb{C}^{N\times N}$ is the forward cyclic shift matrix and 
\begin{equation}
    \boldsymbol{\Delta}_{f^\text{d}_r} = \operatorname{diag}\{e^{-j2\pi n f^\text{d}_r T_s}\}_{n=0}^{N-1} = \operatorname{diag}\{e^{-j2\pi n\nu_r/N}\}_{n=0}^{N-1}, 
\end{equation}
while $\boldsymbol{\Gamma}_{\text{CPP}_r} \in \mathbb{C}^{N\times N}$ is the CPP matrix which is defined as
\begin{equation}
\begin{aligned}
    \boldsymbol{\Gamma}_{\text{CPP}_r} = \operatorname{diag} \Big( &
\begin{cases} 
e^{-j2\pi c_1 (N^2 - 2N(\tau_r - n))}, & n < \tau_r, \\ 
1, & n \geq \tau_r,
\end{cases} 
\Big).
\end{aligned}
\end{equation}
When $2N c_1$ is an integer and $N$ is even, the chirp period prefix becomes equivalent to the cyclic prefix  $\boldsymbol{\Gamma}_{\text{CPP}_r} = \mathbf{I}$.

At the receiver, AFDM demodulation is performed on the received signal $\mathbf{r}$ as $\mathbf{y}[m] = \sum_{n=0}^{N-1}\mathbf{r}[n]\Phi_n^*[m],$
where $\Phi^*$ and $\Phi$ form a discrete affine Fourier transform pair. The matrix representation of the demodulated signal can be written as
\begin{equation}
    \mathbf{y} = \mathbf{U}\mathbf{r}
      = \sum_{r=1}^R \beta_r \mathbf{U} \boldsymbol{\Gamma}_{\text{CPP}_r} \boldsymbol{\Delta}_{f^\text{d}_r} \boldsymbol{\Pi}^{\tau_r} \mathbf{U}^H \mathbf{x} + \tilde{\mathbf{w}},
\end{equation}
where $\tilde{\mathbf{w}} = \mathbf{U}\mathbf{w}$. Since $\mathbf{U}$ is unitary, the statistical properties of $\mathbf{w}$ remain unchanged. Define the effective channel matrix for the $r$-th path as $\mathbf{H}_{\text{eff}_r} = \beta_r\mathbf{U} \boldsymbol{\Gamma}_{\text{CPP}_r} \boldsymbol{\Delta}_{f^\text{d}_r} \boldsymbol{\Pi}^{\tau_r} \mathbf{U}^H$. To clearly illustrate the relationship between the modulated symbols $\mathbf{x}$ and the demodulated symbols $\mathbf{y}$, $\mathbf{H}_{\text{eff}_r}$ can be expressed as
\begin{equation}
    \mathbf{H}_{\text{eff}_r}[p, q] = \frac{1}{N} e^{j \frac{2\pi}{N} \big( Nc_1 \tau_r^2 - q\tau_r + Nc_2 (q^2 - p^2) \big)} \mathcal{F}_r(p, q),
    \label{AFDMChannel}
\end{equation}
where $\mathcal{F}_r(p, q) = \sum_{n=0}^{N-1} e^{-j \frac{2\pi}{N} \big( (p-q+\nu_r+2Nc_1\tau_r)n \big)}$ reflects the impact of delay and frequency shift on the channel matrix. Defining $loc_r = (\alpha_r + 2Nc_1\tau_r)_N$, where $(\cdot)_N$ denotes modulo $N$. Moreover, the following property can be derived from (\ref{AFDMChannel})
\begin{equation}
\begin{aligned}
|\mathbf{H}_{\text{eff}_r}[p, q]| &=\left|\frac{1}{N} e^{j \frac{2\pi}{N} \big( Nc_1 \tau_r^2 - q\tau_r + Nc_2 (q^2 - p^2) \big)} \mathcal{F}_r(p, q)\right|\\
&\quad = \left|  \mathcal{F}_r(p, q)/{N} \right|
= \left| {\sin(N\Theta)}/{N \sin(\Theta)} \right|,    
\end{aligned}
\end{equation}
with $\Theta = \frac{\pi}{N}(p-q+loc_r+a_r)$. For fractional Doppler $a_r\neq0$, it can be observed that $\mathcal{F}_r(p, q)$ attains its maximum magnitude at around $q = (p + loc_r)_N$ and decreases as $q$ moves away from $loc_r$ \cite{ref4}. This provides mathematical support for AFDM systems to achieve signal separation in the delay-Doppler domain. Considering the steering vectors of the receive and transmit antennas in MIMO system, the received signal at the $n_r$-th receiving antenna can be expressed as
\begin{equation}
    \mathbf{Y}_{n_r} = \sum_{r=1}^{R} [\mathbf{a}(\mathbf{r};\phi_r)]_{n_r} (\mathbf{H}_{\text{eff}_r} \mathbf{x})\mathbf{a}(\mathbf{t};\theta_r)^T + \tilde{\mathbf{W}}_{n_r}.
\end{equation}
Stacking $\mathcal{Y}(n_r,:,:)=\mathbf{Y}_{n_r}$ gives the received tensor
\begin{equation}
    \mathcal{Y} = \mathcal{Z} + \mathcal{W} \in \mathbb{C}^{N_r\times N \times N_t},
\end{equation}
where $\mathcal{W}$ denotes the corresponding noise in the  {DAFT} domain, and $\mathcal{Z}$ is the noise‑free signal component. Its CP decomposition is given by
\begin{equation}
    \mathcal{Z} = [\![ \mathbf{A}, \mathbf{B}, \mathbf{C} ]\!]
= \sum_r \mathbf{a}(\mathbf{r};\phi_r) \circ \mathbf{b}(\tau_r, \nu_r) \circ \mathbf{a}(\mathbf{t};\theta_r),
\label{yinzijuzhen}
\end{equation}
where $\circ $ denotes the outer product, and 
\begin{equation}
    \mathbf{b}(\tau_r, \nu_r) =  \mathbf{H}_{\text{eff}_r} \mathbf{x} =  \mathbf{U}\boldsymbol{\Gamma}_{\text{CPP}_r}\boldsymbol{\Delta}_{f^\text{d}_r}\boldsymbol{\Pi}^{\tau_r}\mathbf{U}^H\mathbf{x}.
    \label{bbb}
\end{equation}
Based on the above definition, the matrices 
\begin{align}
    &\mathbf{A}= \left[ \mathbf{a}(\mathbf{r};\phi_1),\cdots,\mathbf{a}(\mathbf{r};\phi_R) \right],\\
    &\mathbf{B} = \left[ \mathbf{b}(\tau_1, \nu_1),\cdots,\mathbf{b}(\tau_R, \nu_R) \right],\\
    &\mathbf{C} = \left[ \mathbf{a}(\mathbf{t};\theta_1),\cdots,\mathbf{a}(\mathbf{t};\theta_R) \right]
\end{align}
are the factor matrices of the tensor signal. These factor matrices carry the structural information of the high-dimensional received signal, with each of them containing physical parameters corresponding to different dimensions. By extracting these factor matrices via tensor decomposition, the loss of structural information caused by dimensionality flattening is avoided.

\section{Constrained Tensor Decomposition-Based Target Sensing Algorithm}
This section outlines a constrained tensor decomposition-based algorithm for parameter estimation. First, we propose a manifold-constrained alternating least squares (ALS) algorithm to estimate factor matrices as specified in (\ref{yinzijuzhen}). Subsequently, the relevant sensing parameters of interest are extracted from the factor matrices which are yielded by the preceding steps. The problem of factor matrices extraction for tensor signals can be formulated as
\begin{equation}
\min_{\hat{\mathbf{a}}_r,\hat{\mathbf{b}}_r,\hat{\mathbf{c}}_r}\Big\|\mathcal{Y}-\sum_{r=1}^R \hat{\mathbf{a}}_r\circ\hat{\mathbf{b}}_r\circ\hat{\mathbf{c}}_r\Big\|^2_F .
\label{fenjiewenti}
\end{equation}
The ALS scheme stands as one of the most well-established approaches for solving (\ref{fenjiewenti}), operating by iteratively minimizing the data fitting error through sequential optimization of each factor matrix while holding the other two fixed, until convergence is attained \cite{ref13}. Nevertheless, ALS treats the factor matrices as unstructured random variables rather than leveraging their inherent structural properties, thereby overlooking valuable prior information and consequently undermining both estimation accuracy and algorithmic efficiency. 

To overcome the above drawbacks, and based on the manifold knowledge of the sparse non-uniform array, we introduce an additional constraint into (\ref{fenjiewenti}) and thereby propose the CALS algorithm. After introducing this constraint, problem (\ref{fenjiewenti}) can be accordingly transformed into
\begin{equation}
    \min_{\hat{\mathbf{A}},\hat{\mathbf{B}},\hat{\mathbf{C}}} \Big\|[\![\hat{\mathbf{A}},\hat{\mathbf{B}},\hat{\mathbf{C}}]\!] - \mathcal{Y}\Big\|_F^2, \quad \text{s.t. } \hat{\mathbf{A}} \in \Gamma_1, \hat{\mathbf{C}} \in \Gamma_2,
    \label{youyueshufenji}
\end{equation}
where $\Gamma_1$ denotes the set of matrices whose column vectors have the structure shown in (\ref{rdaoxiangshiliang}) with an additional scaling factor, and the definition of $\Gamma_2$ is similar to that of $\Gamma_1$. The optimization of (\ref{youyueshufenji}) can be achieved via the ALS framework within \( N_\text{iter} \) iterations until convergence. The \( k \)-th iteration of ALS includes three subproblems with respect to variables \( \mathbf{A}, \mathbf{B}\) and \( \mathbf{C} \), respectively. The \( \mathbf{A} \) factor matrix subproblem can be formulated as a constrained  {least-squares (LS)} problem as
\begin{equation}
\begin{aligned}
    \hat{\mathbf{A}}^{(k+1)} \!= \!\arg\! \min_{\hat{\mathbf{A}}} \| \hat{\mathbf{A}}(\mathbf{C}^{(k)} \odot \mathbf{B}^{(k)})^T \!\! -\!\! \mathbf{Y}_{(1)} \|_F^2, \ \text{s.t. } \hat{\mathbf{A}} \in \Gamma_1,
    \label{AALS}   
\end{aligned}
\end{equation}
where $\mathbf{Y}_{(i)}$ denotes the mode-$i$ unfolding of tensor $\mathcal{Y}$. It is challenging to obtain the closed-form and optimal solution to (\ref{AALS}) due to the nonconvex constraint. Alternatively, we obtain its approximate solution in two steps. In the first step, an intermediate variable \(\tilde{\mathbf{A}}^{(k+1)}\) is obtained via an LS problem (\ref{aputongALS}) without considering the constraint \(\mathbf{A} \in \Gamma_1\). Next, This constraint is considered to refine \(\tilde{\mathbf{A}}^{(k+1)}\) via (\ref{ayueshuALS})
\begin{numcases}{}
\tilde{\mathbf{A}}^{(k+1)}= \arg \min_{\mathbf{A}} \| \mathbf{A}(\mathbf{C}^{(k)} \odot \mathbf{B}^{(k)}) ^T - \mathbf{Y}_{(1)} \|_F^2, \label{aputongALS}\\
\mathbf{A}^{(k+1)} = \arg \min_{\mathbf{A}} \| \tilde{\mathbf{A}}^{(k+1)} - \mathbf{A} \|_F^2 \quad \text{s.t. } \mathbf{A} \in \Gamma_1.\label{ayueshuALS}
\end{numcases}
Problem (\ref{ayueshuALS}) can be solved column-by-column via a correlation projection as $[\mathbf{A}^{(k+1)}]_{:,r} = \lambda_r^{k+1}\mathbf{a}(\mathbf{r};\hat{\phi}_r^{k+1})$, where $[\cdot]_{:,r}$ denotes the $r$-th column. Let $\tilde{\mathbf{a}}_r=[\tilde{\mathbf{A}}^{(k+1)}]_{:,r}$ and $\mathcal{I}_\phi$ denote the AoA search interval. The angle and  scale factor are 
\begin{equation*}
\hat{\phi}_r^{k+1} = \arg\max_{\phi\in\mathcal{I}_\phi} \frac{|\mathbf{a}(\mathbf{r};\phi)^H\tilde{\mathbf{a}}_r|^2}{\|\mathbf{a}(\mathbf{r};\phi)\|_2^2},
\lambda_r^{k+1}= \frac{\mathbf{a}(\mathbf{r};\hat{\phi}_r^{k+1})^H\tilde{\mathbf{a}}_r}{\|\mathbf{a}(\mathbf{r};\hat{\phi}_r^{k+1})\|_2^2}.
\end{equation*}
After obtaining $\hat{\mathbf{A}}^{(k+1)}$, the factor matrix $\hat{\mathbf{B}}^{(k+1)}$ can be obtained by solving a LS problem with $\mathbf{A}$ and $\mathbf{C}$ fixed
\begin{equation}
\hat{\mathbf{B}}^{(k+1)}=\arg\min_{\hat {\mathbf{B}}}\Big\|\mathbf{Y}_{(2)}-\hat{\mathbf{B}}(\mathbf{C}^{(k)} \odot\mathbf{A}^{(k+1)})^T\Big\|_F^2,
\end{equation}
and the factor matrix $\hat{\mathbf{C}}^{(k+1)}$ can be obtained by solving a problem analogous to (\ref{youyueshufenji})
\begin{equation}
    \hat{\mathbf{C}}^{(k+1)}\!\! =\! \arg\min_{\hat{\mathbf{C}}\in\Gamma_2}
    \|\hat{\mathbf{C}}(\mathbf{B}^{(k+1)}\odot\mathbf{A}^{(k+1)})^T-\mathbf{Y}_{(3)}\|_F^2.
    \label{CALS}
\end{equation}

{ 
The identifiability of the factor matrices follows from the essential uniqueness of the CP decomposition. Let $k_{\mathbf X}$ denote the largest integer such that every subset of $k_{\mathbf X}$ columns of $\mathbf X$ is linearly independent. Kruskal's sufficient condition is\cite{ref10}
\begin{equation}
 k_{\mathbf A}+k_{\mathbf B}+k_{\mathbf C}\geq 2R+2.
 \label{cp_uniqueness}
\end{equation}
When (\ref{cp_uniqueness}) holds, any alternative rank-$R$ factorization satisfies $\widetilde{\mathbf A}=\mathbf A \mathbf{P} \boldsymbol\Lambda_A$, $\widetilde{\mathbf B}=\mathbf B\mathbf{P}\boldsymbol\Lambda_B$, and $\widetilde{\mathbf C}=\mathbf C\mathbf{P}\boldsymbol\Lambda_C$, where $\mathbf{P}$ is a common permutation matrix and the nonsingular diagonal matrices satisfy $\boldsymbol\Lambda_A\boldsymbol\Lambda_B\boldsymbol\Lambda_C=\mathbf I$. Thus, each recovered column corresponds to the same target, with its complex scale distributed among the three factors. Because AFDM modulation can separate the received signals in the DAFT domain, it is reasonable to assume that the received signals from different paths are linearly independent under the premise of a proper design of the transmitted signals.  Moreover, sparse antenna arrays provide a larger array aperture and finer angular resolution. These properties establish  favorable conditions for a larger Kruskal rank of the factor matrices $\mathbf{A}$, $\mathbf{B}$ and $\mathbf{C}$.
 
}

Following the recovery of the factor matrices, we process the extraction of the sensing parameters. It should be pointed out that solving problems (\ref{AALS}) and (\ref{CALS}) yields the estimates of the angle parameters $\theta$ and $\phi$. The extraction of delay and Doppler parameters from the complex AFDM signal model is sophisticated. Based on the pulse compression principle, we first adopt a parameter estimation scheme analogous to matched filtering to perform coarse estimation of the integer delay. Recall the vector representation in (\ref{bbb}), the \( q \)-th entry of \( \mathbf{b}(\tau_r, \nu_r) \) can be expressed as
\begin{equation}
    \begin{aligned}
    \mathbf{b}_r[q] = \frac{\beta_r}{N} \sum_{q'=0}^{N-1} \mathbf{x}[q'] &e^{j2\pi\left( -\frac{\tau_r q'}{N} + c_1 \tau_r^2 + c_2 (q'^2 - q^2) \right)}\\
    \times &\sum_{p=0}^{N-1} e^{j\frac{2\pi p}{N} (q' - q - loc_r)}\mathcal{F}_r[q, q'].
    \label{bb}
    \end{aligned}
\end{equation}
Since \( c_2 \) is usually set to be a number sufficiently smaller than  
\((1/2N_c)\) \cite{ref12}, the value of \( N_c c_2 (q'^2 - q^2) \) is approximately  
approach zero, which can be rationally ignored. It is worth noting that  
\( F_r[q, q'] \) in (\ref{bb}) can be simplified to  
$\frac{e^{j \frac{2\pi}{N} (q' - q - loc_r)}}{e^{j \frac{2\pi}{N} (q' - q - loc_r)} - 1}$. For pulse compression, define $\ell_q=\lfloor q/(2Nc_1)\rfloor$ and $j_{m,q}=(m+q)_N$. The coarse correlation function is given as
\begin{equation}
u_r[q] = \sum_{m=0}^{N-1} \hat{\mathbf{b}}_r^*[m]\mathbf{x}[j_{m,q}]
e^{-j2\pi\ell_q j_{m,q}/N}.
\end{equation}
Let $\widehat{loc}_r=\arg\max_q|u_r[q]|$, retaining only the delay indices satisfying $0\leq\ell_q\leq\tau_{\max}$. For $0\leq\alpha_r<2Nc_1$ and $2Nc_1\tau_{\max}+\lceil\nu_{\max}\rceil<N$, the non-wrapped location relation can be utilized to derive the coarse estimates
\begin{equation}
    \hat{\tau}_r = \left\lfloor \frac{\widehat{loc}_r}{2N c_1} \right\rfloor, \quad \hat{\alpha}_r = \widehat{loc}_r-2N c_1 \hat{\tau}_r. 
    \label{zsshiyanpinyi}
\end{equation}

The coarse estimates can be further used to initialize an iterative one dimensional search for delay and Doppler. Here, integer delays can be treated as a special case of fractional delays. Under the stated  conditions and notations, we have
\begin{equation}
 (\hat\tau_r,\hat\nu_r)=\arg\max_{\ell\in\mathcal L,\,g\in\mathcal I_r} f(\ell,g)=
 \frac{|(\boldsymbol\Delta_g\boldsymbol\Pi^\ell\mathbf s)^H\mathbf U^H\hat{\mathbf b}_r|^2}{\|\mathbf s\|_2^2},
 \label{f}
\end{equation}
with $\mathcal I_r=[\hat\alpha_r-1,\hat\alpha_r+1]\cap[0,\nu_{\max}]$ and  $\mathcal L$ follows the similar definition, which represent reasonable intervals for parameter selection. The algorithm alternately optimizes $g$ with $\ell$ fixed and then optimizes $\ell$ with $g$ fixed, thereby gradually approaching the parameter combination that maximizes the objective function $f$. For each delay candidate, a golden-section search contracts the Doppler bracket $[a,b]=\mathcal I_r$ using $g_1=b-\eta(b-a)$ and $g_2=a+\eta(b-a)$, where $\eta=(\sqrt5-1)/2$ represents the interval reduction factor. Specifically, if $f(\ell,g_1)\leq f(\ell,g_2)$, the lower bound is updated as $a=g_1$. Otherwise, the upper bound is updated as $b=g_2$, thereby narrowing the search interval for the next iteration. The interval is successively reduced until its width reaches the residual threshold or $I_\nu$ iterations have been completed.

{\AddToHook{shipout/before}{\ifnum\value{page}>5\relax\PackageError{WCL}{Paper exceeds five pages}{Shorten the three revised passages.}\fi}

Next, we conduct a theoretical analysis of the proposed algorithm. The computational complexity of the proposed two-stage algorithm is composed of that of the CALS stage and the parameter estimation stage. Specifically, the complexity of the CALS stage is given by $\mathcal{O}\{N_{\text{iter}}(3N_tN_rNR+R^2(N_tN+N_rN+N_tN_r)+RG_{\text{ang}}(N_t+N_r))\}$, where $G_{\text{ang}}$ denotes the number of angular grid points used in solving subproblem~(\ref{ayueshuALS}). The dominant term in the complexity of the parameter estimation stage is $\mathcal{O}\{RN^2+RI_\nu N\}$ for direct pulse compression and refinement, after the $\mathcal{O}\{RN\log N\}$ inverse DAFT preprocessing. Then, we further presents the brief derivation of the Cram\'er-Rao bound (CRB), the theoretical limit for the variance of unbiased estimator, which is used to assess the estimation results. For ease of calculation, we use the continuous delay interpolation $\mathbf P(\tau_r) = \mathbf F^H\operatorname{diag}\{e^{-j2\pi k\tau/N}\}_{k=0}^{N-1}\mathbf F$. Define $\mathbf g_r=\mathbf U\boldsymbol\Delta_{\nu_r}\mathbf P(\tau_r)\mathbf U^H\mathbf x$ and $\mathbf v_r=\mathbf a(\mathbf t;\theta_r)\otimes\mathbf g_r\otimes\mathbf a(\mathbf r;\phi_r)$. The observation mean is $\boldsymbol\mu=\mathbf V\boldsymbol\beta$, $\mathbf V=[\mathbf v_1,\ldots,\mathbf v_R]$. The  real parameter vector is
\begin{equation}
 \mathbf p=[\boldsymbol\tau^T,\boldsymbol\nu^T,\boldsymbol\theta^T,\boldsymbol\phi^T,
 \Re\{\boldsymbol\beta\}^T,\Im\{\boldsymbol\beta\}^T]^T.
\end{equation}
For complex Gaussian noise with known variance $\sigma^2$, the log-likelihood equals $L=-\|\operatorname{vec}(\mathcal Y)-\boldsymbol\mu\|_2^2/\sigma^2$ up to a parameter-independent constant. Let $\mathbf D=\partial\boldsymbol\mu/\partial\mathbf p^T$ and its columns follow by differentiating the temporal and spatial factors, with gain derivatives $\mathbf v_r$ and $j\mathbf v_r$. The joint information matrix $\mathbf{J}$ and covariance bound CRB can be given as
\begin{equation}
 \mathbf J=\frac{2}{\sigma^2}\Re\{\mathbf D^H\mathbf D\},
 \qquad \operatorname{CRB}(\mathbf p)=\mathbf J^{-1}.
 \label{conditional_crb}
\end{equation}
For nonsingular matrix \(\mathbf{J}\), the diagonal entries of \(\mathbf{J}^{-1}\) yield lower bounds on the variances of locally unbiased estimators. 
}

\section{Numerical Simulations}
This section provides numerical simulations to assess the accuracy of the proposed algorithm.  We consider a scenario where an ISAC base station is equipped with $N_t = \text{8}$ and $N_r = \text{8}$ linear sparse non-uniform antenna arrays, whose maximum normalized antenna aperture is set to $\text{20}$. The system carrier frequency is $f_c = \text{28}\,$GHz, the transmission bandwidth is $B = \text{100}\,$MHz, and it is divided into $N = \text{128}$ chirp subcarriers. The number of targets is set to $R = \text{5}$, and the target parameters to be estimated are randomly generated within reasonable ranges. The maximum normalized delay and Doppler shift are $\tau_{\max} = \text{8}$ and $\nu_{\max} = \text{2}$, respectively. AFDM parameters can be set with $c_1 = \text{7/256}$ and $c_2 = \text{0}$. Doppler refinement uses $\varepsilon_\nu=\text{10}^{-\text{6}}$ as the residual threshold and at most $I_\nu=10$ golden-section iterations. Multiple signal classification (MUSIC) algorithm and ALS with no constrain are selected to be the baseline, and we choose the normalized mean square error (NMSE) as the metric for evaluating the estimation performance, defined as $\text{NMSE}_\mathbf{X} = \frac{\left\| \mathbf{X} - \hat{\mathbf{X}} \right\|_F^2}{\left\| \mathbf{X} \right\|_F^2}$.

\begin{figure}[H]
    \centering
    \begin{tabular}{@{}c@{}c@{}}
        \begin{subfigure}[b]{0.5\linewidth}
            \centering
            \includegraphics[width=\linewidth]{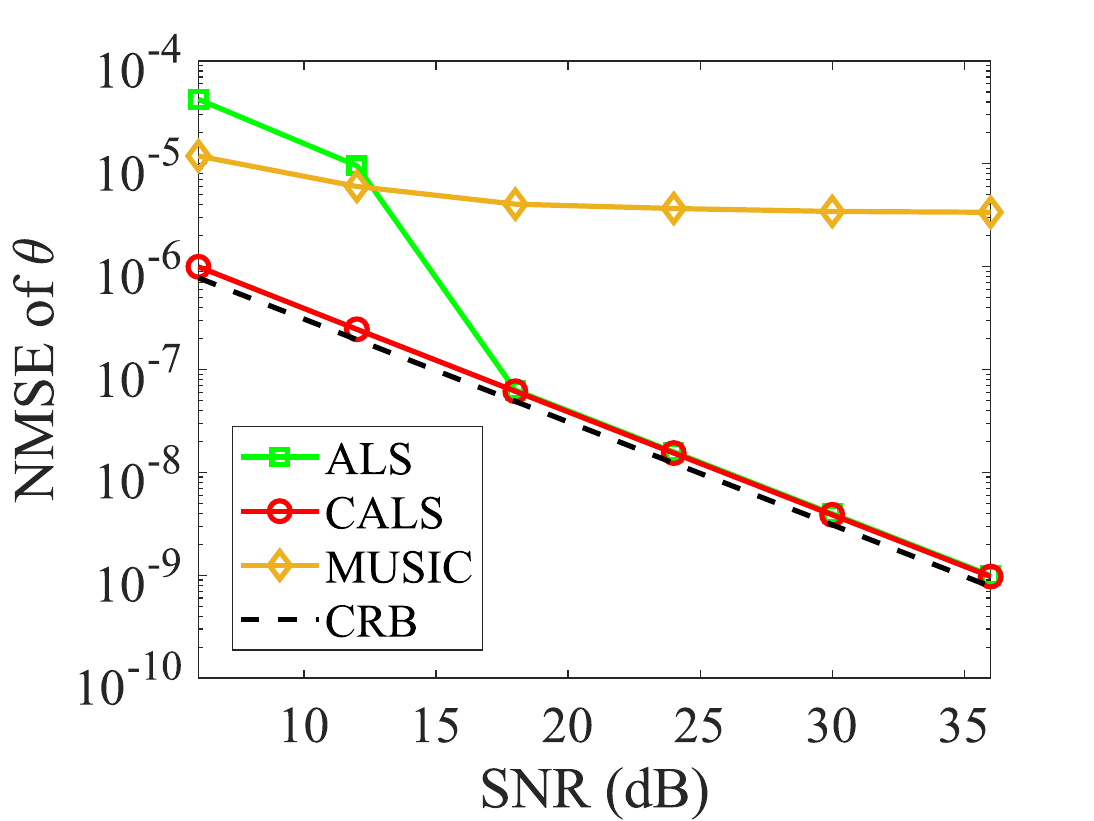}
            \label{theta}
        \end{subfigure}
        &
        \begin{subfigure}[b]{0.5\linewidth}
            \centering
            \includegraphics[width=\linewidth]{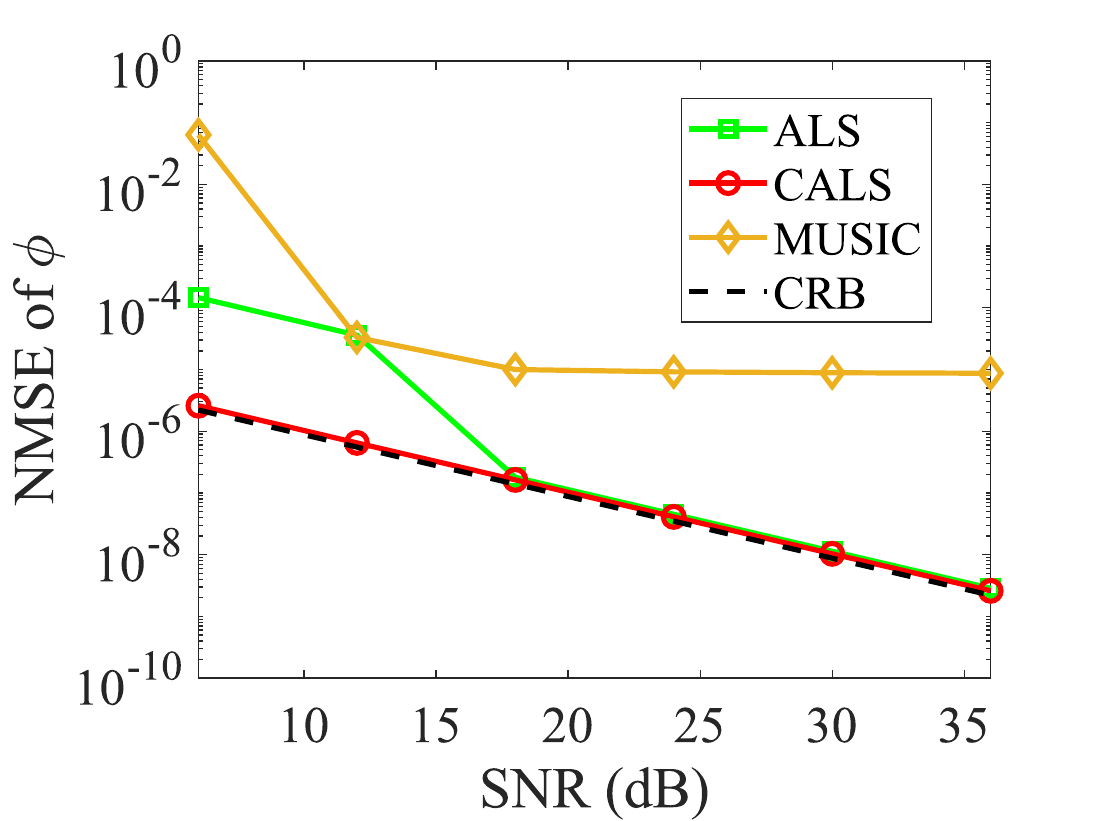}
            \label{phi}
        \end{subfigure}   
\end{tabular}
\caption{NMSE of the AoD (left) and AoA (right).}
\label{angular_nmse}
\end{figure}
First, the estimation performance of the proposed algorithm for the angular parameters of the sensing targets is evaluated. It is observed that the proposed algorithm achieves the best estimation performance and approaches the CRB. Compared with the conventional unconstrained ALS, the proposed algorithm exhibits relatively higher estimation accuracy, with a more pronounced advantage particularly in the low signal-to-noise ratio regime. This is attributed to the fact that the proposed algorithm exploits the position information of the sparse non-uniform antenna arrays in the spatial domain. By utilizing the antenna manifold as a constraint, the iteration is rendered more stable, which enables the factor matrices $\mathbf{A}$ and $\mathbf{C}$ to be decomposed more accurately toward their true values. Consequently, the overall constrained tensor decomposition framework becomes more precise, thereby leading to superior estimation accuracy.

\begin{figure}[H]
    \centering
    \begin{tabular}{@{}c@{}c@{}}
        \begin{subfigure}[b]{0.5\linewidth}
            \centering
            \includegraphics[width=\linewidth]{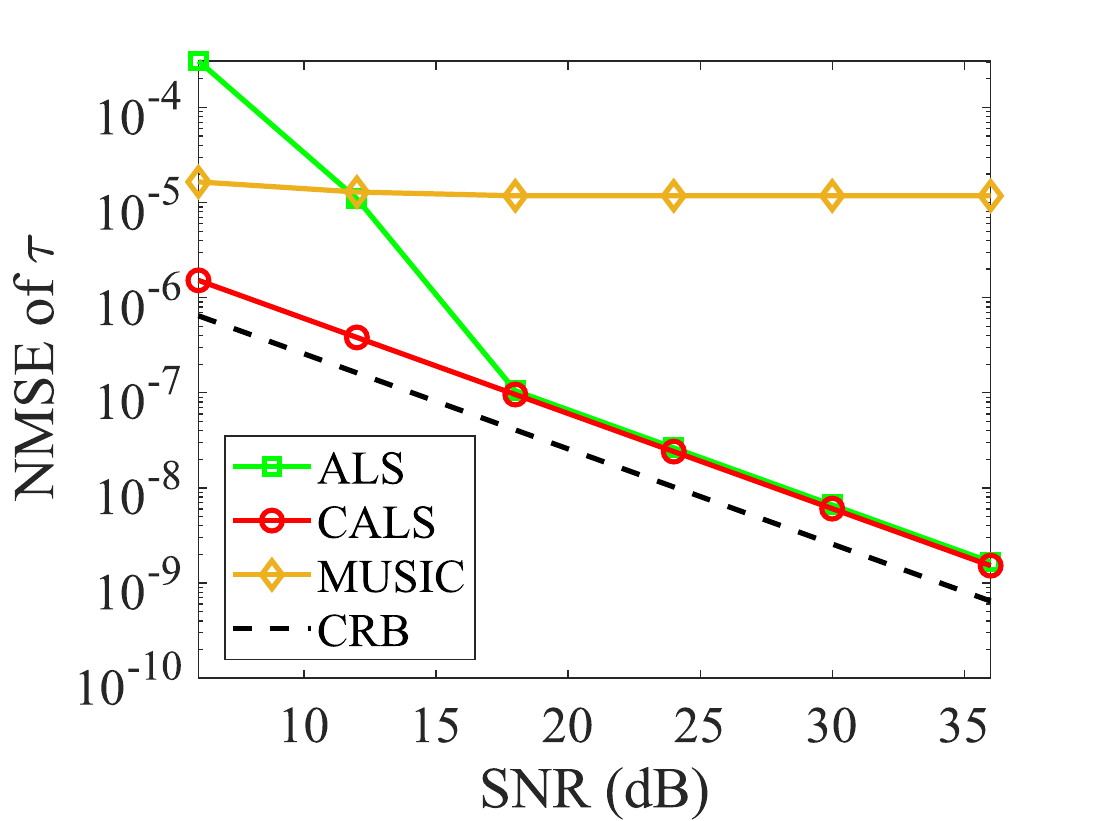}
            \label{tau_nmse}
        \end{subfigure}
        &
        \begin{subfigure}[b]{0.5\linewidth}
            \centering
            \includegraphics[width=\linewidth]{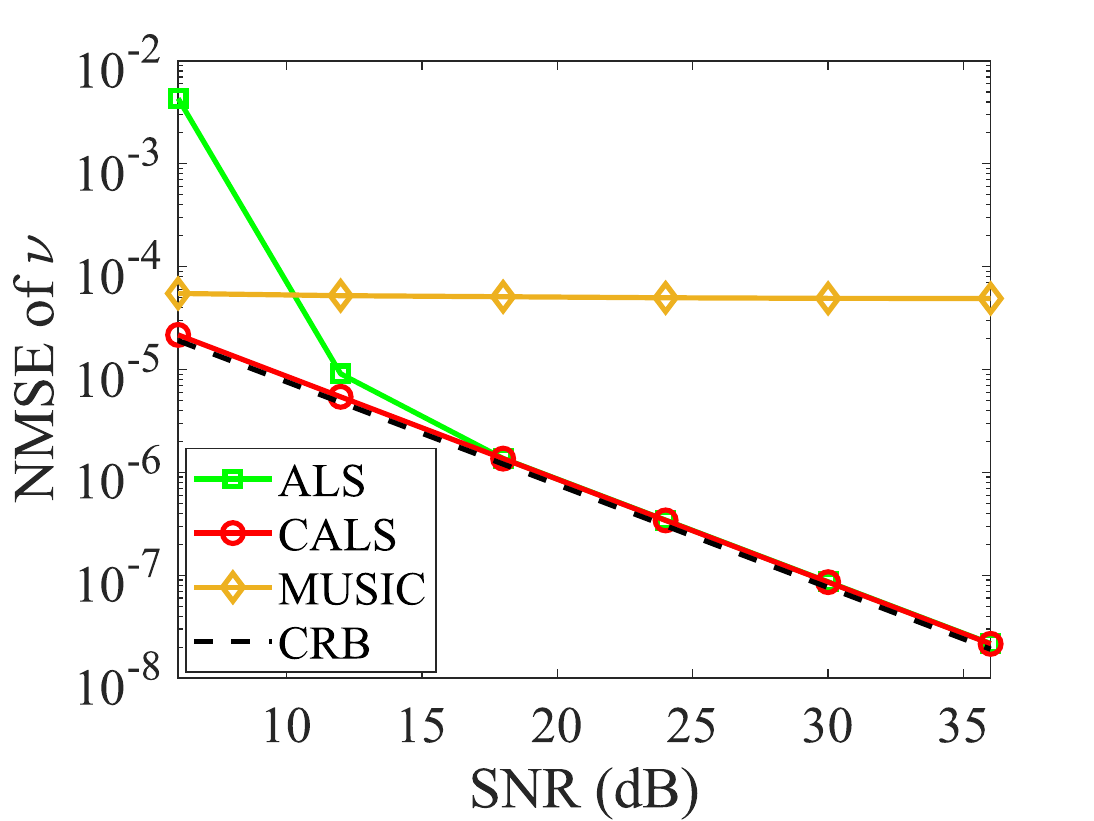}
            \label{nu_nmse}
        \end{subfigure}   
\end{tabular}
\caption{NMSE of time delay (left) and Doppler shift (right).}
\label{delay_doppler_nmse}
\end{figure}
Then, we evaluate the estimation performance for the normalized delay and Doppler shift parameters of the received signal. These two parameters correspond to the position and velocity of the sensing targets, and are of particular importance in AFDM systems. As can be observed from the figure, the proposed algorithm still achieves the best estimation accuracy. Moreover, both tensor decomposition based estimation algorithms outperform the conventional MUSIC algorithm. This is because MUSIC fails to exploit the high-dimensional structural information inherent in the received signal. Instead, it simply flattens the received signal into a matrix for parameter extraction, thereby discarding the structural information contained in the factor matrices. Consequently, its estimation accuracy is inferior to that of the proposed algorithm.

\section{Conclusion}
In this paper, we developed a constrained tensor decomposition based sensing framework for delay, Doppler, and angle estimation in sparse non-uniform array-enabled AFDM ISAC systems. To fully exploit the inherent multidimensional structure of the received echo signals, a tensor model was first established by jointly characterizing the time, frequency, and spatial domains. Based on this model, a manifold-constrained ALS algorithm was proposed to explicitly incorporate the sparse array geometry into the tensor decomposition, enabling robust factor matrix extraction and direct angle estimation. The delay and Doppler shifts were first coarsely estimated via pulse compression and subsequently refined through an iterative one dimensional golden section search using the decomposed factor matrices. Finally, the effectiveness of the proposed algorithm is validated through simulation experiments. Future work will focus on extending the proposed framework to more complex propagation environments, as well as validating its  accuracy and robustness in more practical ISAC systems.

\bibliographystyle{IEEEtran}
\bibliography{ref}

\vfill

\end{document}